\documentstyle[aps,prl,multicol,eqsecnum]{revtex}
\begin{document}

\title{Clustering of inertial particles in turbulent flows}
\author{E. Balkovsky$^{a,c}$, G. Falkovich$^{b,c}$, and A. Fouxon$^{b,c}$}
\address{$^{a}$ School of Mathematics, Institute for Advanced Study,
Princeton, NJ 08540 USA\\ $^{b}$ Physics of Complex Systems, Weizmann
Institute of Science, Rehovot 76100, Israel\\ $^{c}$ Institute for
Theoretical Physics, UCSB, Santa Barbara, CA 93106, USA}
\maketitle

\begin{abstract}
We consider inertial particles suspended in an incompressible
turbulent flow. Due to inertia of particles, their velocity field
acquires small compressible component. Its presence leads to a new
qualitative effect --- possibility of clustering. We show that this
effect is significant for heavy particles, leading to strong
fluctuations of the concentration.
\end{abstract}  

\begin{multicols}{2}

\section{Introduction}

Observing air bubbles in water or dust in air, one readily notices
that inertial particles suspended in an inhomogeneous flow tend to
cluster. For example, such clustering is widely used for flow
visualization. Here we develop a statistical theory of this
phenomenon. We describe the initial growth of concentration
fluctuations and the saturation of that growth due to finite-size
effects, imposed either by the Brownian motion or finite distance
between the particles. Such theory is supposed to have numerous
geophysical and astrophysical applications. A proper account of
concentration fluctuations is also necessary for a consistent theory
of turbulent suspensions.

Macroscopic description of a dilute suspension can be deduced from the
the behavior of a single particle. Consider a small spherical particle
with the radius $a$ and the material density $\rho_0$ suspended in a
fluid with the density $\rho$. The particle's velocity ${\bbox v}$ is
related to the fluid velocity ${\bbox u}$ by the equation $d{\bbox
v}/dt-\beta d{\bbox u}/dt=({\bbox u}-{\bbox v})/\tau_s$, where
$\beta=3\rho/(\rho+2\rho_0)$ and $\tau_s=a^2/(3\nu\beta)$ is the
Stokes time \cite{Maxey}. Both ${\bbox v}$ and ${\bbox u}$ are
evaluated on the particle's trajectory ${\bbox q}(t,{\bbox r})$ that
satisfies $\partial_t{\bbox q}\!=\!{\bbox v}$ and ${\bbox q}(0,{\bbox
r})\!=\!{\bbox r}$. The flow surrounding the particle is assumed to be
viscous, which requires $a\ll r_{\rm v}$, where $ r_{\rm v}$ is the
viscous scale of the flow. This allows one to solve the system for
${\bbox v}$ and ${\bbox q}$ perturbatively in $\tau_s$
\begin{equation}
{\bbox v}={\bbox u}+(\beta-1)\tau_s[\partial_t{\bbox u}+
({\bbox u}\cdot\nabla){\bbox u}]\,.
\label{uv}
\end{equation}
If such particles are spatially distributed, it is possible to define
the particles' velocity field ${\bbox v}(t,{\bbox r})$, which is
compressible even if the fluid flow is incompressible
\cite{Maxey,EKR96}: $(\nabla\cdot{\bbox v})=(\beta-1)\tau_s
\nabla[({\bbox u}\cdot\nabla){\bbox u}]$. Thus in the above expansion
we keep the terms up to the first term with non-vanishing divergence.
As we show the smallness of this term may be compensated by large
parameters, so that it can lead to significant effects.

The concentration of the particles satisfies the diffusion-advection
equation
\begin{equation}
\partial_t n+\nabla({\bbox v} n)=\kappa \nabla^2 n\,.
\label{nden}\end{equation}
Every particle produces a relative perturbation of the flow that
decays as an inverse distance from the particle, i.e. as $a/r$. Since
particles move coherently within the viscous scale $ r_{\rm v}$, the
condition $a\int_a^{ r_{\rm v} }n(r)r^{-1}d^3r \simeq na r_{\rm v}
^2\ll 1$ has to be satisfied in order to be able to neglect their
interaction. This condition is more restrictive than that of a small
concentration, $na^3\ll 1$. If $na r_{\rm v} ^2\ll 1$, the
concentration field can be considered as passive, i.e. ${\bbox v}$ is
independent of $n$ in Eq. (\ref{nden}).

In statistically steady flows, velocity ${\bbox v}$ in
Eq. (\ref{nden}) must be considered as a random field with a
stationary statistics. Evolution of an arbitrary initial condition
$n(0,{\bbox r})$ according to Eq. (\ref{nden}) ultimately results in
the steady state of the concentration fluctuations. Making the
decomposition $n(0,{\bbox r})=n_0+\delta n({\bbox r})$, where $n_0$ is
the spatial average of $n(0,{\bbox r})$, one can write the solution in
the form
\begin{eqnarray}&&
n(t,{\bbox r})\!=\!n_0\!\int \!d{\bbox r}'\, G(t, {\bbox r}, {\bbox r}')
\!+\!\int \!d{\bbox r}'\, G(t,{\bbox r},{\bbox r}')\delta n({\bbox r}')
\label{decompose}\end{eqnarray}
We note that the Green's function, $G$, is nonnegative. At $\kappa=0$
it is concentrated on the Lagrangian trajectory that passes through
the observation point, ${\bbox r}$, at time $t$. At non-zero $\kappa$,
the Green's function is non-zero in some region around the
Lagrangian trajectory. The size of that support region grows in time due to
the combined action of the velocity and diffusion. As long as that size is
smaller than the correlation length of the concentration, $\delta n$
can be taken out of the
integral. Then the expectation value of $n$ depends on the ratio of
$n_0$ and the strength of initial fluctuations $\delta n$. At large
times when the support of the Green's function becomes much larger than
the correlation length, the second term contains contributions that cancel each
other. The expectation value of $n$ is then determined by the first
term. Even though the second term in Eq. (\ref{decompose}) may grow at
these times, it is much smaller than the first term and only gives
subleading dependencies. Thus the initial inhomogeneities of the
concentration field are irrelevant in studying the long-time evolution
and the steady state. Therefore, we will consider a uniform
initial concentration below. We will show that evolution of the uniform
initial concentration distribution exemplifies most strikingly the
inadequacy of the mean field picture for the problem. We choose the
units so that $n_0=1$.

The paper is organized as follows. First, we analyze the initial
period of growth when one can set $\kappa=0$ (to be referred to as the
ideal case). We show that at this stage of evolution the moments of
the concentration grow exponentially for quite an arbitrary velocity
statistics (in case of time-decorrelated velocity this has been shown
in \cite{KG99}). The next section is devoted to the analysis of larger
times. We show that diffusion modifies the growth of the moments and
finally brings about saturation.

\section{The Ideal case}\label{IDEAL}

The diffusivity of macroscopic particles is usually very small so that
we will be interested in the statistics of $n$ in the limit of small
$\kappa$. Starting with a uniform distribution one expects that at
moderate times the diffusion term can be neglected until very thin
structures are developed. Let us consider this period of evolution.
 
To find the concentration $n(t_1,{\bbox r}_1)$ one has to count all
the particles that come to a small volume (still, containing many
particles) around ${\bbox r}={\bbox r}_1$ and divide the result by
this volume. To find which particles comes to this volume one has to
track the trajectories of the particles backwards in time to $t=0$.
Particles perform combined Lagrangian and Brownian motions. At
$t\approx t_1$, the size $l$ of the region occupied by particles grows
as $l^2(t)\sim \kappa (t_1-t)$. Velocity gradient $\lambda$ produces
another mechanism of size stretching which takes over when
$l\gtrsim\sqrt{\kappa/\lambda}$. Smallness of diffusion means that the
Schmidt number, ${\rm Sc}\equiv\nu/\kappa$ is large so that
$\sqrt{\kappa/\lambda}\ll r_{\rm v}$. As one goes further backwards in
time the impact of diffusion on particles' motion becomes negligible.
The time needed for $l$ to reach the diffusive scale
$\sqrt{\kappa/\lambda}$ is of the order of $1/\lambda$. On a larger
time-scale the role of diffusion is to create an initial volume of
finite size $\sqrt{\kappa/\lambda}$. In other words, diffusion
introduces the smallest scale into the problem, so that fluid
particles cannot be localized on the distances smaller than this
scale. To find the concentration at a point in the ideal case one
should track back an infinitesimal volume around that point. However,
in the case of a nonzero diffusivity one should take the volume with a
finite size $\sqrt{\kappa/\lambda}$. Since $\sqrt{\kappa/\lambda}$ can
be smaller than the minimal scale of coarse-graining $n^{-1/3}$ we
introduce $r_{\rm d}$, which is given by the largest of two scales:
$\sqrt{\kappa/\lambda}$ and $n^{-1/3}$. To summarize, $n(t_1,{\bbox
r}_1)$ is given by the relative change of the volume of the size
$r_{\rm d}$ taken around the point ${\bbox r}_1$ and tracked backwards
in time along the Lagrangian trajectory. Let us note that the velocity
gradient fluctuates in a random flow and so does the diffusion
scale. We neglect these fluctuations (c.f. \cite{CKV97}) because they
do not change the dependence of the concentration on large parameters,
which are either time in the transient regime or Reynolds and Schmidt
numbers in the steady state.

At $t\lesssim\lambda^{-1}\ln( r_{\rm v}/r_{\rm d})$ the $r_{\rm
d}$-volume (the region that acquires the size $r_{\rm d}$ at $t=t_1$)
always stays within the viscous scale and hence evolves in a uniform
velocity gradient. Its relative change is the same as for an
infinitesimal volume, which means that the concentration behaves the
same as in the ideal case. Equation (\ref{nden}) written in the
Lagrangian frame therefore becomes the ordinary differential equation
$dn/dt=-n(\nabla\cdot{\bbox v})$. Here $(\nabla\cdot{\bbox v})$ is a
function of time, which fluctuates in a random flow. If the Lagrangian
correlation time of the fluid velocity, ${\bbox u}$, is finite, which
is true for most flows of geophysical interest, then
$(\nabla\cdot{\bbox v})\propto\nabla[({\bbox u}\cdot\nabla){\bbox u}]$
has also a finite correlation time, $\tau$. At $t\gg\tau$, the
concentration logarithm, $X(t)\equiv\ln\left[
n(t)/n(0)\right]=-\int_0^t(\nabla\cdot{\bbox v})dt'$, is a sum of a
large number of random variables. The theory of large deviations
assures that the probability density function (PDF) has the form
${\cal P}(X)\propto \exp[-ts(X/t)]$, where $s$ is a non-negative
convex function \cite{Ellis}. To calculate the moments of the
concentration in the Eulerian frame one has to take every Lagrangian
element with its own weight proportional to its volume, i.e. to the
inverse concentration (see Eqs. (\ref{otcof}) and (\ref{ella})
below). We obtain
\begin{eqnarray}
\langle n^\alpha(t,{\bbox r})\rangle\propto
\int dX\exp[(\alpha-1) X-ts(X/t)]\,.
\end{eqnarray}
At large times, this integral can be found using the saddle-point
approximation. The saddle-point $X_\alpha$ is given by
$s'(X_\alpha/t)=\alpha-1$, which implies $X_\alpha\propto t$. Hence
the moments generally behave exponentially in time: $\langle
n^\alpha(t)\rangle\propto \exp(-\gamma(\alpha) t)$.

Let us show that the conclusion on exponential behavior of moments is
enough to establish the most interesting properties of this stage of
evolution. The number of particles is conserved, i.e. $\left\langle
n\right\rangle$ is time- independent. Hence $\gamma(1)=0$. It is also
obvious that $\gamma(0)=1$. Due to the H\"older inequality, the
function $\gamma(\alpha)$ is convex. Therefore, $\gamma(\alpha)$ is
positive for $0<\alpha<1$ and negative otherwise. Low-order moments
decay whereas high-order and negative moments grow. The decay rate is
$\langle\log |n|\rangle/t=-d\gamma(\alpha)/d\alpha|_{\alpha=0}<0$,
i.e. $n$ decays almost everywhere. Since the mean concentration is
conserved, $n$ has to grow in some (smaller and smaller) regions,
which implies growth of high moments. The growth of passive scalar
fluctuations in the particular case of a short-correlated compressible
flow was described in \cite{KG99}.

The finite value of the coarse-grained volume comes into play at
$t\gtrsim\lambda^{-1}\ln(r_{\rm v}/r_{\rm d})$. Indeed, since
particles separates backwards in time, the size of the volume exceeds
$ r_{\rm v}$ at $t=0$. In other words, spots originated from different
viscous domains come into contact at $t=t_1$. A careful analysis of
the time-span of the ideal case approximation demands more detailed
information on the Lagrangian dynamics at scales smaller than $ r_{\rm
v}$ (later referred to as small-scale Lagrangian dynamics). It is
discussed in subsection \ref{ssld}.

\subsection{Small-scale Lagrangian dynamics of compressible fluids
with finite Lagrangian correlation time}
\label{ssld}

Let us present general analysis of Lagrangian statistics assuming
that the distances between particles are smaller than $r_{\rm v}$.
Consider two Lagrangian trajectories ${\bbox q}(t,{\bbox r}_1)$ and
${\bbox q}(t,{\bbox r}_2)$, satisfying the equations $\partial_t{\bbox
q}(t,{\bbox r}_i)={\bbox v}(t,{\bbox q}(t,{\bbox r}_i)) $ and the
initial conditions ${{\bbox q}}(0, {\bbox r}_i)={\bbox r}_i$.  The
distance between the two trajectories ${\bbox R}={\bbox q}(t, \bbox
r_1)-{\bbox q}(t, {\bbox r}_2)$ satisfies the equation $\partial_t\bbox
R={\bbox v}(t,{\bbox q}(t, {\bbox r}_1))- {\bbox v}(t,{\bbox q}(t, \bbox
r_2))\approx \sigma {\bbox R}$, when $R$ is much smaller than the
viscous length and the velocity difference can be approximated by the
first term in its Taylor expansion. The rate-of-strain matrix,
\begin{eqnarray}&&
\sigma_{\alpha\beta}(t)=\left.\frac{{\partial v_{\alpha}}(t,
{\bbox r})}{ {\partial r_{\beta}}}\right|_{{\bbox r}={\bbox q}(t,
\bbox r_2)}
\end{eqnarray}
is what determines the deformation of a fluid blob of a small size. If
the velocity statistics is spatially homogeneous, then the uniform
sweeping is statistically irrelevant and only the deformation part of
the Lagrangian transformation is significant. It is given by ${\bbox
R}=W {\bbox R_0}$, where the evolution matrix $W$ satisfies
\begin{eqnarray*}&&
\partial_t  W=\sigma  W\,,\quad
\left.  W\right|_{t=0}= 1\ .
\end{eqnarray*}
The Lagrangian evolution may be thus considered as a linear mapping
given by the affine transformation $ W$, whose statistics is
determined by the statistics of $\sigma$.  Such Lagrangian mapping can
be described universally at times much larger than the correlation
time, $\tau$, of $\sigma$. In the one-dimensional case one finds $\ln
W\equiv\rho=\int_{0}^t dt' \sigma(t)$, so that at $t\gg\tau$ one deals
with a sum of large number of independent random variables. The
probability distribution function (PDF) of $\rho$ is completely
determined by the entropy function $S$ (see e.g. Ref. \cite{Ellis})
\begin{eqnarray}&&
{\cal P}(t,\rho)=
\frac{1}{Z(t)}\exp\left[-tS\left(\frac{\rho-\lambda t}{t}\right)\right],
\end{eqnarray}
where $\lambda$ is the average value $\langle\sigma\rangle$ and $1/Z$
is the normalization factor. At $\rho=\lambda t$ the PDF has a sharp
maximum, described by the central limit theorem, where $S(x)\approx
x^2/(2C)$. Here $C$ is the dispersion of $\sigma$:
\begin{eqnarray*}&&
C=\int dt'\langle\langle\sigma(t)\sigma(t')\rangle \rangle\,.
\end{eqnarray*} 

This analysis can be generalized to higher dimensions. The idea is
briefly presented here following the recent exposition in
\cite{BF99}.  At large times, the Lagrangian transformation can be
represented as a stretching or contraction along fixed orthogonal
directions followed by a rotation. Indeed, one can represent $ W$ as
the product $ M \Lambda N$, where $ M$ and $ N$ are orthogonal
matrices, and $ \Lambda$ is a diagonal matrix
\cite{Goldhirsh,GCS96}. At large times the matrix $ N$ becomes
asymptotically constant, as follows from the Oseledec theorem for $
W^t W$ \cite{Osel}. Excluding the constant matrix $ N$ by the proper
choice of initial basis one finds $ W= M
\Lambda$. This representation shows that in the frame rotating with
the fluid blob the Lagrangian mapping is just a stretching along fixed
directions.  Stretching accumulates so that the characteristic time of
change of the matrix $\Lambda$ is $t$ while the matrix $M$ changes on
the much shorter time-scale which is of order of the inverse elements
of $\sigma$. Since the time scales of $ M$ and $ \Lambda$ are widely
different, the matrices are statistically independent. Indeed,
changing $\sigma$ at the last stage of evolution with duration $\tau$
one changes $M$ completely, whereas $\Lambda$ is changed by the amount
of the order $\tau/t\ll 1$. It means that fixing the value of
$\Lambda(t)$ does not change the distribution of $M(t)$. The matrix
$M$ is uniformly distributed over the rotation group. The PDF of the
eigen values of the matrix
$\Lambda=diag\left[\exp(\rho_1),\ldots,\exp(\rho_d)\right]$, is given
by
\begin{eqnarray}&&
{\cal P}(t,\rho_i)=
\frac{1}{Z}\exp\left[-tS\left(\frac{\rho_1-\lambda_1t}{t},\ldots,
\frac{\rho_d-\lambda_dt}{t}\right)\right]\nonumber\\&&
\times\theta(\rho_1-\rho_2)...\theta(\rho_{d-1}
-\rho_d)\,.
\label{prob}
\end{eqnarray}
Here $\theta$ is the step function that orders the eigen values, so
that $\rho_1\geq\rho_2\geq..\geq\rho_d$. The constants $\lambda_i$
(ordered in the same way) are the Lyapunov exponents of the flow. In
principle, they can be expressed via the statistics of $\sigma$ (see
e.g. \cite{BF99,Goldhirsh}). We will assume that the Lyapunov spectrum
is non-degenerate, i.e. $\lambda_1>\lambda_2>\ldots>\lambda_d$. The
normalization factor $Z$ in Eq. (\ref{prob}) is a function of
time. Equation (\ref{prob}) shows that at times much larger than the
correlation time of $\sigma$, independently of the statistics of
$\sigma$, the statistics of $ W$ is characterized by a single function
$S$ of $d$ variables. This entropy function is convex and positive,
and it has the expansion
\begin{eqnarray}&&
S=x_i C^{-1}_{ij}x_j
\label{nima}\end{eqnarray}
near its minimum at $x_1=\ldots=x_d=0$. Here $C_{ij}$ is the
covariance matrix of $\sigma$. Note that we assume the entropy
function to be nonzero at least in some interval of $\rho_i$ which
physically means that the flow is random. We also assume that the flow
is compressible so that $S$ is a regular function of all variables
(the distribution in the incompressible case is obtained by a limiting
operation). The homogeneous dependence of the PDF on $\rho_i$ and $t$
is often sufficient to establish universal statistical laws
independently of the details of velocity statistics
\cite{BF99,CFKV99,BFL99}.

At large times, ${\cal P}(t,\rho_i)$ has a sharp maximum at
$\rho_i=\lambda_i t$. The existence of generally non-zero Lyapunov
exponents manifests itself in various growth rates. For example, the
average logarithmic rate of separation of two particles located within
the viscous scale of the flow is given by $\lambda_1$, the
corresponding growth rate of the fluid volume is $\sum_{i=1}^d
\lambda_i$ etc.

For an incompressible random flow $\lambda_1>0$ \cite{Zeldovich},
while the average value of $\sigma$ is zero $\langle\sigma\rangle=0$.
The appearance of non-zero $\lambda_1$ is related to the interplay
between rotational and stretching degrees of freedom
\cite{95CFKL,BF99,Zeldovich} (for a scalar equation one would have 
$\lambda_1=\langle \sigma\rangle=0$). The simplest way to appreciate
the existence of positive Lyapunov exponent is to consider (following
\cite{Zeldovich}) an example of a saddle-point 2d flow $v_x=\lambda
x,v_y=-\lambda y$ where the stretching directions satisfy
$\cos\phi\geq(1+\lambda^2)^{-1/2}$ that is their measure is larger
than $1/2$.

Compressibility introduces another mechanism of correlations affecting
the stretching.  There are more Lagrangian particles in the
contracting regions with $\nabla\cdot {\bbox v}<0$, leading to to the
appearance of negative average gradients in the Lagrangian frame. By
isotropy one has $d\langle
\sigma_{\alpha\beta}\rangle=\delta_{\alpha\beta}\langle
\left.\nabla\cdot {\bbox v}(t,{\bbox r})\right|_{{\bbox r}={\bbox
q}(t, {\bbox r}_0)}\rangle$. Averaging over the volume, one obtains
\begin{eqnarray}&&
\langle\left.\nabla\cdot {\bbox v}(t,{\bbox r})\right|_{{\bbox r}={\bbox
q}(t, {\bbox r}_0)}\rangle=\int\frac{d{\bbox r}_0}{V}
\left.\nabla\cdot {\bbox v}(t,{\bbox r})\right|_{{\bbox r}
={\bbox q}(t, {\bbox r}_0)}
\nonumber
\\&&
=\int\frac{d{\bbox r}}{V}\nabla\cdot {\bbox v}(t,{\bbox r})
\left(\left.\det\left|\left|\frac{\partial {\bbox q}(t, {\bbox r}_0)}
{\partial r_0}\right|\right|\right)^{-1}\right|_{{\bbox q}(t, \bbox
r_0)={\bbox r}}
\nonumber
\\&&
=\left\langle \nabla\cdot {\bbox v}(t,{\bbox r})
\left(\left.\det\left|\left|\frac{\partial {\bbox q}(t, {\bbox r}_0)}
{\partial r_0}\right|\right|\right)^{-1}\right|_{\bbox q(t, \bbox
r_0)={\bbox r}}\right\rangle,
\label{compr}
\end{eqnarray}
We observe that this Lagrangian average generally coincides with the
Eulerian average $\int d{\bbox r} \nabla\cdot{\bbox v}(t,{\bbox r})/V$
only in the incompressible case, where it is zero. In the compressible
flow, the integral (\ref{compr}) is zero at zero time (when we set the
initial conditions for the Lagrangian trajectories so the measure is
uniform back then). The average is getting time-independent and
negative at times larger than the velocity correlation time when the
compressed regions with negative ${\bbox
\nabla}\cdot{\bbox v}$ acquire higher weight than the expanded ones. Let us
illustrate this conclusion by considering the physically interesting
case of $\nabla\cdot {\bbox v}$ short-correlated in
time (to be discussed in much detail below). 
Taking $t$ in (\ref{compr}) larger than the correlation
time of $\nabla\cdot {\bbox v}$ yet small enough to allow for the
expansion
\begin{eqnarray*}
{\det}^{-1}\left|\left|\frac{\partial{\bbox q}( t, {\bbox r})}
{\partial{\bbox r}}\right|\right|\approx 1-\int_0^{ t}\nabla
\cdot {\bbox v}(t',{\bbox r})dt'\,,
\end{eqnarray*}
one finds
\begin{eqnarray*}&&
\langle
\left.\nabla\cdot {\bbox v}(t,{\bbox r})\right|_{{\bbox r}={\bbox
q}(t, {\bbox r}_0)}\rangle=\nonumber\\&&-\frac{1}{2}
\int_0^t dt'\langle \nabla\cdot {\bbox v}(t,{\bbox r})
\nabla\cdot {\bbox v}(t',{\bbox r})\rangle\,.
\end{eqnarray*}

Negative $\langle tr\sigma\rangle=\nabla\cdot {\bbox v}$ leads to a
suppression of the stretching by the velocity field. If one decomposes
$ \sigma$ into ``incompressible and compressible parts''
\begin{eqnarray*}&&
 \sigma_{\alpha\beta}=
\left( \sigma_{\alpha\beta}-\frac{1}{d}\delta_{\alpha\beta}\,tr\sigma
\right)+
\frac{1}{d}\delta_{\alpha\beta}\,tr\sigma\,,
\end{eqnarray*}
then from the explicit expressions for Lyapunov exponents
\cite{BF99,Goldhirsh} it is easy to find that the Lyapunov exponents
of the incompressible process $\sigma_{\alpha\beta}-
\delta_{\alpha\beta}tr\sigma/d$ (with $\lambda_1>0$) get uniformly
shifted down by $\langle tr\sigma\rangle/d$. At a sufficient degree of
compressibility, all the exponents may become negative (in the
one-dimensional case where compressibility is maximal one can prove
that $\lambda<0$, see below).

Let us stress the difference in Eulerian and Lagrangian averages
appearing in the compressible case. An Eulerian average is uniform
over the space while in a Lagrangian average every trajectory comes
with its own weight determined by the local rate of volume change.
This difference is of an utmost importance in the discussion of
backward in time Lagrangian statistics to which we pass now.

We have seen that to find the concentration we must consider the
evolution of a fluid blob backwards rather than forward in time. This
is a general situation: to find the value of an advected field at the
given space-time point, ${\bbox r},T$, one should consider the
Lagrangian trajectory ${\bbox q}(t| T, {\bbox r})$ fixed by its final
(rather than initial) position:
\begin{eqnarray*}&&
\partial_t {\bbox q}(t| T, {\bbox r})=
{\bbox v}(t,{\bbox q}(t| T, {\bbox r})),\ \ 
{\bbox q}(T| T, {\bbox r})={\bbox r}.
\end{eqnarray*}
The initial point ${\bbox q}(0| T, {\bbox r})$ depends on the velocity
realization, so that it is random and not fixed as in the above
analysis. We denote the Lagrangian quantities related to the
trajectories fixed by their destination by the tilde sign. The strain
matrix is defined as
\begin{eqnarray}&&
\tilde\sigma_{\alpha\beta}=\left.
\frac{\partial v_{\alpha}(t,{\bbox r})}{\partial r_\beta}\right|_{{\bbox r}=
q(t| T, {\bbox r}_0)}\,.
\end{eqnarray}
This must be compared with the definition of $\sigma$, where the
initial condition fixes the Lagrangian trajectory. The matrices
$\sigma$ and $\tilde\sigma$ generally have different statistical
properties. The evolution matrix $\tilde W$ is now defined by
\begin{eqnarray}&&
\partial_t \tilde W(t| T, {\bbox r})=\tilde \sigma \tilde W(t| T, 
{\bbox r}),\ \ \tilde W(0| T, {\bbox r})=1\,.
\nonumber\end{eqnarray}
Its value at $t=T$ determines the deformation of fluid blobs coming to
a fixed final point. Let us relate $\tilde W$ to $W$ which we now
write with a spatial argument
\begin{eqnarray*}&&
W_{ij}(t| t', {\bbox r})=\frac{\partial q_i(t| t', {\bbox r})}
{\partial r_j}.
\end{eqnarray*}

The expression for $\tilde W(t| T, {\bbox r})$ in terms of $W(t| t',
{\bbox r})$ is given by $\tilde W(t| T, {\bbox r})=W(t| T, \bbox
r)W^{-1}(0| T, {\bbox r}) $, so that $\tilde W(T)=W^{-1}(0| T, \bbox
r_0)$. Differentiating the identity ${\bbox q}(T| 0, {\bbox q}(0| T,
{\bbox r}))={\bbox r}$ one finds $W(T| 0, {\bbox q}(0| T, {\bbox r}))=
W^{-1}(0| T, {\bbox r})$, which relates $W$ and $\tilde W$: 
\begin{eqnarray*}&&
\tilde W(T| T, {\bbox r})= W(T| 0, {\bbox q}(0| T, {\bbox r}))\,.
\end{eqnarray*}

To express the statistics of $\tilde W$ in terms of the Lagrangian
characteristics we use the same transformation that we used for
transforming the average of $\nabla\cdot {\bbox v}$ in the Lagrangian
frame to the usual Eulerian average. Namely, for the average over the
volume of the flow of any function $f$ one has
\begin{eqnarray}&&
\langle f\{\tilde{W}(T)\}\rangle=\int\frac{d{\bbox r}_0}{V}
f\{W(T|0, {\bbox q}(0| T, {\bbox r}_0))\}\nonumber\\&&=\int\frac{d\bbox
x}{V} f\{W(T| 0, {\bbox x})\}\det\left|\left|\frac{\partial {\bbox q}(T| 0,
{\bbox x})} {\partial x}\right|\right|
\nonumber
\\&&
=\left\langle f\left\{W(T| 0, {\bbox x})\right\}\det W(T| 0, \bbox
x)\right\rangle\,.
\label{avgs}\end{eqnarray}
Again, Lagrangian and Eulerian averages coincide for incompressible
flow, when $\det W\equiv1$. In general, it is necessary to account for
the local volume change when passing from one average to
another. Since $\det W=\exp\sum\rho_i$, then considering passive
fields one has to take the following probability distribution of
stretching/contraction eigen values
\begin{eqnarray}&&
\tilde{\cal P}(t,\rho_i)\!=\!\frac{1}{Z}\exp\left[\sum_{i=1}^d\rho_i\!-\!
tS\left(\frac{\rho_1-\lambda_1t}{t},\ldots,
\frac{\rho_d-\lambda_dt}{t}\right)\right]\nonumber\\&&
\times\theta(\rho_1-\rho_2)\ldots\theta(\rho_{d-1}
-\rho_d)\ .\label{prob1}
\end{eqnarray}
Here $S$ and $Z$ are the same as in (\ref{prob}) that is what one can
measure in studying (forward-in-time) particle dispersion. Note that
the correct normalization of $\tilde{\cal P}$ is guaranteed by the
volume conservation $\left\langle\det W\right\rangle=1$ following from
(\ref{avgs}) with $f=1$.
 
The Lyapunov exponents $\tilde{\lambda_i}$ of $\tilde{W}$ are
determined by the extremum of the exponent:
\begin{eqnarray}&&
\tilde{\lambda_i}=\lambda_i+y_i,
\label{y_i}
\end{eqnarray}
where $y_i$ are determined from ${\partial S(y_1,..,y_d)}/{\partial
y_i}=1$. An important remark is that $y_i$ cannot generally be
expressed via the Lyapunov exponents $\lambda_i$ only. They depend on
the form of the entropy function $S$ and hence on the details of
velocity statistics. Indeed, every trajectory comes with its own
weight determined by the local rate of volume change. The consequence
is that passive fields behavior in a compressible flow does not enjoy
the same degree of universality as in the incompressible case (when,
for instance, the growth rate of the magnetic fluctuations is
determined solely by the spectrum of the Lyapunov exponents
$\lambda_i$ and is independent of the form of the entropy function
\cite{CFKV99}).

Even though we will use only the properties of the matrix $\tilde W$,
we also mention the probability distribution function of the eigen
values of matrix $W(0| T, {\bbox r})$ which directly determines the
evolution backwards in time
\begin{eqnarray*}&&
\tilde{\cal P}(t,\rho_i)\!=\!\frac{1}{Z}\exp\Biggl[-\sum_{i=1}^d\rho_i\!-\!
tS\Biggl(\frac{-\rho_1-\lambda_dt}{t},\ldots,
\nonumber
\\&&
\frac{-\rho_d-\lambda_1t}{t}\Biggr)\Biggr]
\theta(\rho_1-\rho_2)\ldots\theta(\rho_{d-1}
-\rho_d)\ .
\end{eqnarray*}
which follows from the above results using $W(0| T, {\bbox r})=
\tilde W^{-1}(T| T, {\bbox r})$. It follows that the backward-in-time
Lyapunov exponents are given by $-\tilde\lambda_i$ and not by the
naive guess $-\lambda_i$, which holds only in the incompressible
case. In particular, particles diverge backwards in time with exponent
$-\tilde\lambda_d$.

The difference between Lyapunov exponents $\lambda_i$ and
$\tilde\lambda_i$ can be illustrated in the one-dimensional case where
their signs are definite and opposite. Indeed, the conservation of the
total fluid volume together with spatial homogeneity imply
\begin{eqnarray}&&
\langle W\rangle=\frac{1}{Z}\int d\rho\exp\left[\rho-
tS\left(\frac{\rho-\lambda t}{t}\right)\right]=1.
\label{avdet}\end{eqnarray} 
Let us consider this identity at large times when the integral can be
calculated by the saddle-point method. First, we note that at large
$t$ the normalization factor $Z=\int d\rho \exp(-tS)$ is determined by
the region, where $S(x)\propto x^2$ so that $Z\propto t^{1/2}$. Next,
one can rewrite the integral (\ref{avdet}) as $tZ^{-1}\int
dx\exp[t(\lambda+x-S(x))]$. It is determined by the point $x_*$ where
$x-S(x)$ is maximal. Since $x-S(x)$ takes positive values near $x=0$
we conclude that $x_*-S(x_*)>0$. Therefore $\lambda= S(x_*)-x_*<0$,
for the integral (\ref{avdet}) to be time-independent.

On the contrary, $\tilde \lambda$ is positive. Defining
$\rho+S((\rho-\lambda t)/t)\equiv \tilde S((\rho-\tilde\lambda t)/t)$
one has the condition $\int d\rho \exp[-\rho-t\tilde
S(\rho/t-\tilde\lambda)]=1$ which gives $\tilde\lambda>0$.

The above results are generalized to higher dimensions as follows.
Since we consider the flow to be contained in a fixed volume then
$\langle \det W\rangle=1$, and in the same manner one finds that the
mean logarithmic rate-of-change of the volume elements $\sum
\lambda_i\leq0$ (which, in particular, implies $\lambda_3\leq0$). From
the explicit expressions for $\lambda_i$
\cite{BF99,Goldhirsh} one finds $\sum \lambda_i =\langle\nabla\cdot\bbox
v\rangle$, so that we have proved that the Lagrangian average
$\langle\nabla\cdot{\bbox v}\rangle$ is nonpositive, the result stated
above on physical grounds. The corresponding inequality on
$\tilde\lambda_i$ is $\sum \tilde\lambda_i\geq0$ (implying $\tilde
\lambda_1\geq0$). We arrive at a somewhat surprising conclusion that
for the Lagrangian dynamics one has average compression of volumes,
whereas passive fields rather feel average expansion. The physical
meaning of this effect is transparent: as we go away (either forward
in calculating $\lambda_i$ or backwards in calculating
$\tilde\lambda_i$) from the moment where we imposed a uniform
Lagrangian measure, the volume rate-of change is getting negative in a
fluctuating compressible flow. To avoid misunderstanding, let us
stress that for a physical quantity $x(t)$ (volume of a fluid element
in this case) the conservation of the mean value $\langle x(t)\rangle$
does not contradict to a nonzero rate of change $t^{-1}\langle\ln
x(t)\rangle$.

The general considerations can be illustrated using a particular case
of the velocity statistics, the Kraichnan model of a short-correlated
Gaussian velocity with the variance
\begin{eqnarray}&&
\left\langle v_{\alpha}(t,{\bbox r})v_{\beta}(0,0)\right\rangle
=\left[V_0\delta_{\alpha\beta}-{\cal K}_{\alpha\beta}({\bbox r})\right]
\delta(t)\,,
\label{vvar}
\\&&
{\cal K}_{\alpha\beta}=D\left[(d+1-2\Gamma)
\delta_{\alpha\beta}r^2+2(d\Gamma-1)r_{\alpha}r_{\beta}\right]\,.
\label{short}
\end{eqnarray}
Here $\Gamma$ is the ratio of the variances of $\nabla_\alpha
v_\alpha$ and $|\bbox{\nabla v}|$ respectively, it is thus the degree
of compressibility that may vary between 0 and 1. The quadratic
dependence of the correlation function on the coordinate corresponds
to the expansion of the velocity difference we made above. For a
velocity defined by Eqs. (\ref{vvar}) and (\ref{short}), a
straightforward calculation gives
\begin{equation}\lambda_i/D=d(d+1-2i)-2\Gamma[d+(d-2)i]\ .\label{lambda}
\label{lmbd}\end{equation}
In the incompressible case, $\Gamma=0$, this formula has been derived
in \cite{CGK}.  For a general compressible case, $\lambda_1$ has been
derived in \cite{CKV97}, where it has been also observed that
$\lambda_1$ changes sign at $\Gamma=d/4$. The entropy function has the
form (\ref{nima}) for arbitrary values of $x$. One can also find
\begin{eqnarray*}&&
C_{ij}=4D\{[d+\Gamma(d-2)]\delta_{ij}-1+\Gamma d\}\,.
\end{eqnarray*}
We see from (\ref{lambda}) that compressibility indeed diminishes the
Lyapunov exponents. It is interesting to compare this with the
influence of compressibility on Lagrangian dynamics in a multiscale
velocity: there, the Lagrangian trajectories either explosively
separate or implosively collapse depending on whether the degree of
compressibility is small or large respectively \cite{GV}.

The Lyapunov exponents $\tilde\lambda_i$ that govern the behavior of
the passive fields are enhanced by compressibility since $y_i$ are
positive (see Eq. (\ref{y_i})). For the Kraichnan model one has
$y_i=\sum_j C_{ij}/2$, so that
\begin{equation}
\tilde\lambda_i/D=d(d+1-2i)+2\Gamma[d^2-(d-2)i-2]\ .
\label{tilyambda}
\end{equation}
For $d=2,3$ one has
\begin{eqnarray*}&&
\tilde\lambda_1=2D(1+2\Gamma)\,,\quad
\tilde\lambda_2=-2D(1-2\Gamma)\,,
\\&&
\tilde\lambda_1=6D(1+2\Gamma)\,,\quad
\tilde\lambda_2= 10D\Gamma\,,\quad
\tilde\lambda_3/=-2D(3-4\Gamma)\,.
\end{eqnarray*}
The compressibility, $\Gamma$, is identically equal to unity in $d=1$,
where instead of Eq. (\ref{short}) one should write ${\cal
K}=Dx^2$. Then, Eqs. (\ref{lmbd}) and (\ref{tilyambda}) are replaced
by $\lambda=-D$ and $\tilde\lambda=D$ respectively.

Comparison of (\ref{lmbd}) and (\ref{tilyambda}) shows that
$\lambda_i=-\tilde\lambda_{d+1-i}$. This relation is due to
time-reversibility of the short-correlated velocity. In particular,
$\lambda_1$ and $\tilde\lambda_d$ change sign at the same degree of
compressibility $\Gamma=d/4$. This is peculiar for a short-correlated
case and does not hold for an arbitrary velocity statistics. In other
words, the change of the regime from stretching to contraction in the
forward Lagrangian dynamics does not generally correspond to the
change of the regime in the passive fields, which are related to the
backwards in time Lagrangian dynamics.

\subsection{Applicability of the ideal case approximation}

Let us now determine the domain of validity of the ideal case
approximation. Our starting point will be the dynamic expression for
the Green's function that can be derived explicitly in this case. To
find when the concentration starts to be determined by the particles
that were initially separated by a distance larger than the viscous
scale, we must analyze the support of the Green's function, $G(t,
{\bbox r}| t'=0,{\bbox r}')$, as a function of the initial coordinate,
${\bbox r}'$. At $\kappa=0$ the dynamics is purely Lagrangian, so that
\begin{eqnarray*}
G(t, {\bbox r}| 0, {\bbox r}')=\frac{1}{\det \tilde W(t, {\bbox r})}
\delta({\bbox r}'-{\bbox q}(0|t, {\bbox r}))\,.
\end{eqnarray*}
This formula has a clear meaning: the change of the concentration at a
point is completely determined by the volume compression factor along
the Lagrangian trajectory. Small diffusion is equivalent to adding a
Brownian motion to the velocity. It leads to a smearing of the region
around the Lagrangian trajectory from which particles come to the
observation point. As a function of the initial arguments, the Green's
function satisfies the Hermite-conjugate evolution equation
\begin{eqnarray}&&
\partial_{t'}G+({\bbox v}(t', {\bbox r}'), \nabla_{r'})G=-\kappa
\nabla_{r'}^2 G\,.
\end{eqnarray} 
As long as the support of $G$ is much smaller than the viscous scale,
one can expand the velocity in the vicinity of the Lagrangian
trajectory in the Taylor series. The homogeneous component is excluded
by passing to the moving frame and the first non-trivial term contains
$\tilde{\sigma}$:
\begin{eqnarray}&&
\partial_{t'}G+\tilde\sigma_{\alpha\beta}(t')
r'_\beta \nabla_{\alpha}G=-\kappa
\nabla_{r'}^2 G\,.
\end{eqnarray}
This equation can be solved in the Fourier space
\begin{eqnarray}&&
G(t, {\bbox r}|0, {\bbox r}')=\frac{1}{\det \tilde W(t, {\bbox r})}
\nonumber\\&&
\times\int
\frac{d{\bbox k}}{(2\pi)^d}\exp\left[i{\bbox k}\cdot[{\bbox r}'-
{\bbox q}(0|t, {\bbox r})]
\!-\!\frac{k^tIk}{2}
\right],
\label{Green}
\\&&
I=2\kappa\int_0^t\!\! dt' \,\tilde W^{-1}(t'| t, {\bbox r})\tilde
W^{-1, t} (t'| t, {\bbox r})\,.
\nonumber\end{eqnarray}
The matrix $I$ is the inertia tensor of a patch of particles,
evaluated at $t=0$, provided the patch is a sphere with the center at
the point ${\bbox r}$ at time $t$. The particles perform independent
Brownian motions together with the Lagrangian motion in the same
velocity field (cf. \cite{BF99}). Let us stress that no averaging has
been performed in Eq. (\ref{Green}) and therefore the expression for
$G$ is purely dynamical. We observe that the size of the region which
makes the main contribution to the concentration at a point grows as
the largest eigen value of the matrix I, i.e. the square of the linear
size grows as $\kappa\int_0^t dt'\exp[-2\tilde\rho_d(t')]$. Since the
diffusionless consideration is valid as long as the largest size of
the $r_{\rm d}$-volume is smaller than the viscous scale, the
applicability condition of Eq. (\ref{Green}) is
\begin{equation}
\kappa\int_0^t dt' \exp[-2\tilde\rho_d(t')]\ll r_{\rm v}^2\,.
\label{apcon}\end{equation}

Below, we will refer to the configurations of velocity with decreasing
$\tilde \rho_d(t)$ as the contracting configurations. Indeed, for such
configurations, particles at the observation point are brought
together from larger regions. On the other hand, for the
configurations with increasing $\tilde\rho_d(t)$, the concentration is
determined by the particles initially belonging to the region of the
size of the order $\sqrt{\kappa/\lambda}$. Such configurations can be
called diverging because the particles in the vicinity of ${\bbox
q}(0| t, {\bbox r})$ diverge exponentially.

Formula (\ref{Green}) gives $n(t, {\bbox r})=\int G(t, {\bbox r}| 0,
{\bbox r}')d{\bbox r}'=1/\det \tilde W(t, {\bbox r})$. This is exactly
the same expression as for non-diffusing particles. For the moments of
the concentration one finds
\begin{eqnarray}&&
\langle n^{\alpha}\rangle=\int d{\bbox \rho}_i \exp[-(\alpha-1)\sum \rho_i]
P(t, \rho_i)
\label{otcof}\end{eqnarray}
with $P(t, \rho_i)$ given by Eq. (\ref{prob}). Note that the growth
function $\gamma_{\rm L}(\alpha)$ in the Lagrangian frame is obtained
by a mere shifting of the argument of the Eulerian growth function:
\begin{equation}
\langle n^{\alpha}(q(t,{{\bbox r}}),t)\rangle= \langle
n^{\alpha+1}({\bbox r}, t)\rangle\,.
\label{ella}\end{equation}

Integral (\ref{otcof}) can be calculated using the saddle-point
approximation. The saddle-point value of $\tilde \rho_d$ given by
$-c_{\alpha}t$, where $c_{\alpha}$ is an $\alpha-$dependent
constant. i.e. the condition of applicability is $\kappa\int_0^t
dt'\exp[2c_{\alpha}t']\ll r_{\rm v}^2$. Using convexity of the entropy
one can show that large negative moments have negative $c_{\alpha}$
and therefore the above condition becomes time-independent. This can
be simply seen noting that the averaged quantity $\exp[-\alpha\sum
\tilde\rho_i]$ favors positive $\tilde\rho_i$ at negative
$\alpha$. Therefore, the diffusionless result is always correct for
large negative moments. Diffusion cannot stop the formation of void
regions with few particles inside.

On the other hand, from expression (\ref{otcof}) one can see that for
$\alpha>0$ any growing moment of $n$ must be determined by a positive
$c_{\alpha}$, otherwise $\sum \tilde\rho_i>\tilde
\rho_d>0$. Generally, one can assert the existence of the boundary
$\alpha_{\rm b}$, such that $-\infty<\alpha_{\rm b}<1$. For
$\alpha<\alpha_{\rm b}$, the saddle-point $c_{\alpha}$ is negative and
the corresponding moment behaves as in the diffusionless case for all
times, whereas for $\alpha>\alpha_{\rm b}$ the diffusionless
approximation breaks down at large times. Since the moments with
$\alpha<\alpha_{\rm b}$ are determined by the configurations on which
the $r_{\rm d}$-volume is compressed backwards in time, one expects
that $\alpha_{\rm b}$ is a monotonically increasing function of the
velocity compressibility (as measured by $\sum \lambda_i$). 
For example, in the framework of the
Kraichnan model one has $\alpha_{\rm
b}=(\Gamma-4d)/(2\Gamma(d+2))$. We will refer to the $\alpha_{\rm
b}<0$ case as the weakly compressible case, and $0<\alpha_{\rm b}<1$
as the strongly compressible one. It can be verified that this
corresponds to the cases of $\tilde \lambda_d<0$ and $\tilde
\lambda_d>0$ respectively. The same is valid for an arbitrary velocity
statistics.

In fact at any time $t$ one can consider the contribution of diverging
configurations, so to say, the ``ideal fluid contribution'':
\begin{eqnarray}&&
\langle n^{\alpha}\rangle_{\rm id}
=\int_{\tilde\rho_d>0} d\rho_i \exp\left[-(\alpha-1)
\sum \tilde\rho_i-tS\right]\,.
\label{idin}\end{eqnarray}
Since the smallest size cannot be smaller than $r_{\rm d}$, it is
necessary to introduce here the cutoff at $\tilde\rho_d=0$. It is
clear that $\langle n^{\alpha}\rangle>\langle n^{\alpha}\rangle_{\rm
id}$. Integral (\ref{idin}) has exponential time-dependence, $\langle
n^{\alpha}\rangle_{\rm id}\propto\exp[\gamma_{\rm id}(\alpha)t]$. Note
that due to the constraint $\tilde\rho_d>0$ the growth function
$\gamma_{\rm id}$ is different from $\gamma(\alpha)$. For
$\alpha<\alpha_{\rm b}$ the saddle-point is inside the domain of
integration at all times. On the contrary, for $\alpha>\alpha_{\rm
b}$, integral (\ref{idin}) is determined by the boundary,
$\tilde\rho_d=0$.

In the weakly compressible situation, $\alpha_{\rm b}<0$, one has
$\gamma_{\rm id}(\alpha_{\rm b})=\gamma(\alpha_{\rm b})>0$. From the
continuity of $\gamma_{\rm id}(\alpha)$ we conclude that $\gamma_{\rm
id}(\alpha)>0$ for $\alpha<\alpha'_{\rm b}<0$, where $\alpha'_{\rm b}$
is defined as $\gamma_{\rm id}(\alpha'_{\rm b})=0$. The inequality
$\alpha'_{\rm b}<0$ follows from $\gamma_{\rm id}(\alpha)
<\gamma(\alpha)$ and $\gamma(0)=0$.
Therefore, the moments of the order $\alpha<\alpha'_{\rm b}<0$ satisfy
$\langle n^{\alpha}\rangle>\exp(\gamma'_{\alpha} t)$ with positive
$\gamma'_{\alpha}$ at all times (in fact, asymptotically the equality
holds as mixing configurations can only lead to a growth slower than
exponential, see below). It means that these moments become infinite
in the steady state, which corresponds to the formation of the
power-law asymptotic behavior for the PDF of the concentration near
$n=0$: ${\cal P}(n)\propto n^{-\alpha'_{\rm b}-1}$. Diffusion does
modify the growth of the moments with $\alpha_{\rm b}
<\alpha<\alpha'_{\rm b}$, but the time dependence remains exponential.

In the strongly compressible case, $\alpha_{\rm b}>0$, one can
immediately conclude that the moments with $\alpha<0$ grow
exponentially with the ideal fluid exponents. For $\alpha>0$ the
inequality $\langle n^{\alpha}\rangle >\langle n^{\alpha}\rangle_{\rm
id}$ leads to no interesting conclusions at large times. The
asymptotic behavior of the concentration PDF at $n\to 0$ is ${\cal
P}(n)\propto n^{-1}$.

Let us now consider the moments to which the main contribution is made
by the contracting configurations (i.e. the moments with
$\alpha>1$). The cutoff time $t^*$ for the ideal growth is determined
from the condition $\exp(2c_{\alpha} t^*_{\alpha})\sim (r_{\rm v}/
r_{\rm d})^2$, which gives $t^*_{\alpha}=c^{-1}_\alpha\ln(r_{\rm
v}/r_{\rm d})$. This expression is exact in the limit of large Schmidt
numbers. Note that the cutoff time depends on the order of the
moment. For a quadratic in $\alpha$ entropy (e.g. for the Kraichnan
model), $c_{\alpha}$ is a linear function of $\alpha$. The
steady-state dependence of the moments on Schmidt number can be
estimated from below by $(r_{\rm v}/r_{\rm d}
)^{-\gamma(\alpha)/c_{\alpha}}$. One can expect that the dependence of
the steady-state moments on the Schmidt number is linear for large
$\alpha$, since $\gamma_{\alpha}\sim \alpha c_{\alpha}$. This can be
seen from the saddle-point expression for the moment. The linear
dependence signifies less intermittent tail of the PDF as compared to
the evolution problem.


\section{Saturation of growth due to diffusion}

To analyze the behavior of the moments at larger times one should
distinguish two cases: the $Re\sim 1$ case when the velocity
correlation length $L$ is of the order of $r_{\rm v}$, and the case of
$Re\gg 1$, when $L\gg r_{\rm v}$. In the first case, advection becomes
equivalent to the usual diffusion at the scales larger than $r_{\rm
v}$. The moments get saturated and are given by the corresponding
power of $r_{\rm v}/r_{\rm d}$. In the large $Re$ case the velocity
divergence is correlated at scales much larger than $r_{\rm v}$. The
correlation between different viscous domains decays as a power of the
distance between them. The configurations that coherently bring
together different viscous domains determine the moments at this stage
of evolution.

The moments continue to grow (in a power-law fashion) only for a
particular case when the compressible correction to the velocity has
the same scaling as the incompressible velocity. Since the
compressible part is proportional to $(u\nabla)u$, then $u$ and $v$
have the same scaling only for a smooth velocity, $\delta u\propto r$,
which has been studied in Sec. \ref{IDEAL}. Note that up to
logarithmic corrections this is true for a vorticity 2d cascade as
well. However, for the turbulent velocity in the energy cascade, the
velocity $u$ is non-smooth, hence the compressible part has a
different scaling. For example, the Kolmogorov phenomenology gives
$\delta u\propto r^{1/3}$), so that $\delta v\propto r^{-1/3}$. It
means that the compressibility is most important at small (viscous)
scale so that the growth has to saturate and the level of fluctuations
should not depend on the Reynolds number.

Unfortunately we still lack the formalism to describe Lagrangian
statistics in the inertial interval with the same degree of
universality as in the viscous interval. Nevertheless, to understand
the most essential properties of the concentration fluctuations, one
can use the simplest velocity statistics. We assume that the velocity
is statistically isotropic, Gaussian, and has zero correlation time
\cite{SS00}. The pair-correlation function is given by
\begin{eqnarray}&&
\langle v_{\alpha}(t,{\bbox r},)v_{\beta}(0, 0)\rangle
\!=\!\delta(t)[V_0\delta_{\alpha\beta}-
{\cal K}_{\alpha\beta}({\bbox r})]\,,
\label{cov}\\&&
(d-1){\cal K}_{\alpha\beta}=\left[\frac{(r^{d+1}u)'}{r^d}-c\right]r^2
\delta_{\alpha\beta}
\!-\!\left[\frac{(r^2u)'}{r}-c\right]r_\alpha r_\beta\,.
\nonumber\end{eqnarray}
Note that $c=0$ for incompressible flows.

We will assume that $u$ and $c$ have a regular expansion at $r\ll
r_{\rm v}$, that is $u(r)\approx u(0)+u''(0)r^2/2+\ldots$, and
$c\approx c(0)+c''(0)r^2/2+\ldots$. In the intermediate region $r_{\rm
v} \ll r \ll L$, the functions $u$ and $c$ behave in power-law
manners. However, due to relation (\ref{uv}), the scaling exponents of
$u$ and $c$ are generally different. At large scales, $r\gg L$, the
function $u$ scales as $r^{-2}$. To guarantee that the variance of the
velocity is positive, $c$ must vanish faster than $r^{-2}$ at $r\gg
L$.

A convenient measure of compressibility is given by the ratio
$\epsilon=c/u$. We will denote by $\epsilon_0$ the value of $\epsilon$
for the smooth velocity ($u,c$ constants). Despite all these crude
simplifications the model enables to see the most interesting features
of the growth and is used to illustrate the conclusions we believe to
be model-independent.

\subsection{The two-point correlation function}

In this subsection we study the two-point correlation function of the
concentration, $f({\bbox r})=\langle n(0)n({\bbox r})\rangle$. In the
framework of the model (\ref{cov}) it satisfies the closed equation
\begin{eqnarray}
\partial_t f=\nabla_\alpha\nabla_\beta \left({\cal K}_{\alpha\beta} f
\right)+2\kappa\nabla^2 f.
\label{pair}
\label{bsc}\end{eqnarray}

Let us first show that the dynamics of $f$ is relaxational and then
find the stationary solution to which it converges. For this purpose
we must analyze the spectrum of the differential operator on the
right-hand side of Eq. (\ref{bsc}). The eigen functions must be
regular at small distances. The boundary condition at large distances
follows from the fact that the correlation function tends to a
constant, equal to $1$ ($n_0^2$ in the dimensional units). Therefore
one must require that the eigen functions do not grow at infinity. The
operator has the form of a $d-$dimensional Fokker-Planck operator, so
that one can expect that it has no positive eigen values. Due to
spherical symmetry of $f$ on can rewrite Eq. (\ref{bsc}) in the
spherical coordinates
\begin{eqnarray}&&
\partial_t f\!=\!r^{1-d}\hat{{\cal L}}_{FP}\left(r^{d-1} f\right)\,,\quad 
\\&&
\hat{{\cal L}}_{FP}\!=\!\partial_r\left[r^2u+2\kappa\right]e^{-\Phi}\partial_r
e^{\Phi}\,,
\\&&
e^{\Phi}=\frac{1}{r^{d-1}}\exp
\left[\int \frac{rc(r)\,dr}{r^2u(r)+2\kappa}\right]\,.
\end{eqnarray}
Note that $\hat{{\cal L}}_{FP}$ has the form of a one-dimensional
Fokker-Planck operator. Now we can show that the operator $\hat{{\cal
L}}$ on the right-hand side of Eq. (\ref{bsc}) has no positive eigen
values, i.e. all the eigen functions of the operator satisfy
$\hat{{\cal L}}f_E=-Ef_E$ with $E>0$. Indeed, the evolution operator
becomes proportional to a Laplacian at $r\gg L$. Therefore the
negative energy eigen functions have exponential behavior at infinity.
The boundary condition at infinity ensures that only exponentially
decaying solutions are allowed. To show that for such solutions the
boundary condition at $r=0$ cannot be satisfied, we write the identity
\begin{eqnarray}&&
\int_0^{\infty}\!dr\, f_E\, e^{\Phi}r^{2d-2}\hat{{\cal L}} f_E=-E
\int_0^{\infty}\!dr\,f_E^2\, e^{\Phi}r^{2d-2}
\nonumber
\\&&
=-\int_0^{\infty}\!dr\,\left(r^2u+2\kappa\right)e^{-\Phi}
\left(\partial_r e^{\Phi}r^{d-1}f_E\right)^2<0\,,
\nonumber
\end{eqnarray}
which proves that $E>0$. We used integration by parts, which is
possible only for functions decaying at infinity.

Let us now show that in fact the spectrum is continuous, covering the
interval $[0, \infty)$. For definiteness we assume that $r_{\rm d}\ll
r_{\rm v}\ll L$. Let us consider the equation at $r\ll r_{\rm v}$
where $u=u(0)$ and $c=\epsilon u(0)$:
\begin{eqnarray}
(1+x^{2})f''+
\left[(d+1+\epsilon)x+\frac{d-1}{x}\right]
f'+(d\epsilon+\lambda)f=0\,.
\nonumber\end{eqnarray}
Above we have assumed that the distance is measured in units $r_{\rm
d}$, where $r_{\rm d}^2=2\kappa/u(0)$. We have also introduced
$\lambda=E/u(0)$. The solution of this equation satisfying the correct
boundary conditions is
\begin{eqnarray}&&
f_E(r)=C(E)F\left(\frac{d+\epsilon-\mu}{4}, \frac{d+\epsilon+\mu}{4}, \frac{d}
{2}, -\left(\frac{r}{r_{\rm d}}\right)^2\right),
\nonumber\end{eqnarray}
where $\mu=\sqrt{(d-\epsilon)^2-4\lambda}$. Next, let us consider the
inertial interval. Considering for simplicity $u=Dr^{-\gamma}$ and
$c=\epsilon_\gamma Dr^{-\gamma}$ we find
\begin{eqnarray}
r^2f''+
(d+1-\gamma+\epsilon)r
f'+\left[(d-\gamma)\epsilon+\frac{Er^{\gamma}}{D}\right]f=0\,,
\nonumber\end{eqnarray}
The solution of this equation is
\begin{eqnarray}&&
f=C_1(E)r^{-(d+\epsilon-\gamma)/2}
J_\nu\left(\frac{2\sqrt{E}}{\gamma\sqrt{D}}r^{\gamma/2}\right)
\nonumber
\\&&
+C_2(E)
r^{-(d+\epsilon-\gamma)/2}
N_{\nu}\left(\frac{2\sqrt{E}}{\gamma\sqrt{D}}r^{\gamma/2}\right),
\label{above}\end{eqnarray}
where $\nu=(d-\gamma-\epsilon)/\gamma$. Finally, at $r\gg L$ one can
use the function (\ref{above}) substituting $\gamma=2$, $\epsilon=0$
and $V_0$ instead of $D$
\begin{eqnarray}&&
f=C_3(E)\,r^{1-d/2}
J_{d/2-1}\left(\sqrt{\frac{E}{V_0}}r\right)
\nonumber\\&&
+C_4(E)\,r^{1-d/2}
N_{d/2-1}\left(\sqrt{\frac{E}{V_0}}r\right),
\nonumber\end{eqnarray}
We observe that unlike the case of $E<0$, for $E>0$ there is no
additional restriction on the coefficients of the eigen function. Thus
the matching problem can always be solved. The matching at $r\sim
r_{\rm v}$ fixes the ratio of constants $C_1/C_2$, and the matching at
$r\sim L_v$ fixes the ratio $C_3/C_4$. We conclude that the spectrum
is positive continuous and non-degenerate. In fact, this property
holds for any relation between $r_{\rm d}$ and $r_{\rm v}$.

Thus at large times $f$ must converge to $f_0\equiv f_{\rm st}$. Note
that decay at large times is prohibited by the inequality
$f(t,0)=\langle n^2\rangle>\langle n\rangle^2=1$. Let us now find the
stationary solution which satisfies
\begin{eqnarray}&&
\partial_r\left(\left[r^2u+2\kappa\right]e^{-\Phi}\partial_r
\left(e^{\Phi}r^{d-1}f_{{\rm st}}\right)\right)=0.
\end{eqnarray}
The function must approach the square of the average concentration at
infinity and be regular at zero. It is easy to see that the solution
satisfying these conditions is the ``zero flux'' solution,
proportional to $\exp[-\Phi]r^{1-d}$. It is given by
\begin{eqnarray}&&
f_{{\rm st}}=\exp\left[\int_r^{\infty}
\frac{r'c(r')dr'}{r'^2u(r')+2\kappa}\right]\ .
\label{st}
\end{eqnarray}
The integral in Eq. (\ref{st}) converges, because the function $c$
decays faster than $r^{-2}$ at $r\gg L$. Since $r^2u\sim V_0$ at $r\gg
L$, we obtain the following asymptotic expression at these scales
\begin{eqnarray}&&
f_{{\rm st}}-1\propto\left(\frac{L}{r}\right)^{\alpha}\,,
\nonumber\end{eqnarray}
where we assumed that $c$ decays as $r^{-2-\alpha}$, $\alpha>0$.

The behavior of $f_{\rm st}$ in the inertial interval, $r_{\rm v}\ll
r\ll L$, crucially depends on whether $u$ and $c$ have the same
scaling exponents. If they do, $f_{{\rm st}}(r)$ behaves as a negative
power of $r$. The single point correlation function, $\langle
n^2\rangle$, which can be estimated as $f_{{\rm st}}(r_{\rm d})$, is
then proportional to a positive power of the Reynolds number. If,
however, the velocity is not smooth in the inertial interval, $u$ and
$c$ have different scaling exponents, and Eq. (\ref{st}) shows that
the main growth of $f_{\rm st}$ occurs below the viscous scale. Hence,
$\langle n^2\rangle$ is independent of the Reynolds number. For
example, for the Kolmogorov scaling, the solution (\ref{st}) has the
form $\ln f_{{\rm st}}\propto a^4r^{-4/3} r_{\rm v}^{-8/3}\beta^{-2}$
at $r\ll r_{\rm v}$.

Therefore, fluctuations of the concentration are mainly produced in
the interval of scales $r\lesssim r_{\rm v}$, where the fluid velocity
is smooth (i.e. in the viscous interval or in 2d vorticity
cascade). One can write estimates
\begin{equation}
\langle n^2\rangle\simeq
 \left(\frac{r_{\rm v}}{r_{\rm d}}\right)^{\epsilon_0}\,,\quad
\epsilon_0\simeq\beta^{-2} \left(\frac{a}{r_{\rm v}}\right)^4\,.
\label{est}\end{equation}
Since $r_{\rm d}$ is by definition larger than $a$, significant
fluctuations are possible only for very heavy particles with
$\beta\approx 2\rho/3\rho_p <(a/r_{\rm v})^2\ln^{1/2}(r_{\rm v}/r_{\rm
d})$.

To conclude, the fluctuations of concentration grow exponentially in
any random compressible flow with a nonzero sum of Lyapunov exponents
until this growth is restricted by finite-size effects (diffusion or
discreetness of the particles). It is interesting to note that $n$ can
be considered as the density of the fluid itself, so that the
finite-size effects are absent. Since density perturbations does not
grow unlimited, one concludes that the phenomenon described here can
take place only as a transient process when, for instance, large-Mach
random flow with almost homogeneous density was initially created. In
a stationary turbulence of the compressible fluid, the back reaction
of density fluctuations on the flow stop the growth of the density
fluctuations (we are indebted to V. Lebedev for this remark).

The work has been supported by NEC Research Institute, Minerva
Foundation and Mitchell Research Fund. We thank V. Lebedev and
M. Vergassola for useful discussions.

\end{multicols}

\end{document}